\documentclass[twocolumn,useAMS,usenatbib]{mn2e}
\usepackage{graphicx}
\usepackage{natbib}

\usepackage{bm}
\usepackage{amsfonts}
\usepackage{latexsym}
\usepackage[latin1]{inputenc}
\usepackage{amsmath}
\usepackage{amssymb}
\usepackage{epsfig}
\usepackage{amsbsy}

\newcommand{\n}{\mathrm n}

\newcommand{\p}{\mathrm p}
\newcommand{\e}{\mathrm e}
\newcommand{\x}{\mathrm x}
\newcommand{\y}{\mathrm y}

\newcommand{\mvb}[1]{\boldsymbol{#1}}

\def \nn  {\nonumber}

\def\jnl@style{\rm}
\def\aaref@jnl#1{{\jnl@style#1}}

\def\aaref@jnl#1{{\jnl@style#1}}

\def\aj{\aaref@jnl{AJ}}                   
\def\apj{\aaref@jnl{ApJ}}                 
\def\apjl{\aaref@jnl{ApJ}}                
\def\apjs{\aaref@jnl{ApJS}}               
\def\apss{\aaref@jnl{Ap\&SS}}             
\def\aap{\aaref@jnl{A\&A}}                
\def\aapr{\aaref@jnl{A\&A~Rev.}}          
\def\aaps{\aaref@jnl{A\&AS}}              
\def\mnras{\aaref@jnl{MNRAS}}             
\def\prd{\aaref@jnl{Phys.~Rev.~D}}        
\def\prl{\aaref@jnl{Phys.~Rev.~Lett.}}    
\def\qjras{\aaref@jnl{QJRAS}}             
\def\skytel{\aaref@jnl{S\&T}}             
\def\ssr{\aaref@jnl{Space~Sci.~Rev.}}     
\def\zap{\aaref@jnl{ZAp}}                 
\def\nat{\aaref@jnl{Nature}}              
\def\aplett{\aaref@jnl{Astrophys.~Lett.}} 
\def\apspr{\aaref@jnl{Astrophys.~Space~Phys.~Res.}} 
\def\physrep{\aaref@jnl{Phys.~Rep.}}      
\def\physscr{\aaref@jnl{Phys.~Scr}}       

\title[Magnetic field evolution timescales in NSs]{On the magnetic
  field evolution timescale in superconducting neutron star cores}

\author[A. Passamonti, T. Akg\"un, J.A. Pons, J. A. Miralles] {Andrea
  Passamonti\thanks{E-mail:passamonti@ua.es}, Taner Akg\"un, Jos\'{e}
  A. Pons, Juan A. Miralles\\Departament de F\'{i}sica Aplicada, Universitat d'Alacant,
  Ap. Correus 99, 03080 Alacant, Spain }

\begin{document}

\date{\today}

\pagerange{\pageref{firstpage}--\pageref{lastpage}} \pubyear{}

\maketitle

\label{firstpage}


\begin{abstract}

We revisit the various approximations employed to study the long-term
evolution of the magnetic field in neutron star cores and discuss
their limitations and possible improvements. A recent controversy on
the correct form of the induction equation and the relevant evolution
timescale in superconducting neutron star cores is addressed and
clarified.  We show that this ambiguity in the estimation of
timescales arises as a consequence of nominally large terms that
appear in the induction equation, but which are, in fact, mostly
irrotational. This subtlety leads to a discrepancy by many orders of
magnitude when velocity fields are absent or ignored. Even when
internal velocity fields are accounted for, only the solenoidal part
of the electric field contributes to the induction equation, which can
be substantially smaller than the irrotational part.  We also argue
that stationary velocity fields must be incorporated in the slow
evolution of the magnetic field as the next level of approximation.

\end{abstract}


\section{Introduction} \label{sec:intro}

The evolution of the magnetic field in the interior of neutron stars
is a complex and controversial research area, with several important
open issues. Part of the difficulty stems from the multi-fluid
character of the problem, with at least three different species
(protons, neutrons, and electrons) that are not necessarily coupled in
some range of temperatures and timescales of interest. This makes
necessary a multi-fluid dynamics approach to the problem. In spite of
recent relevant advances in the field, many important aspects are
still under debate. For instance, we still ignore the `equilibrium'
configuration which emerges from the star formation after a
core-collapse, and how the subsequent long-term evolution of the
magnetic field in the core proceeds.

The complex multi-fluid physics becomes particularly difficult when
protons/neutrons in the core undergo a transition to a
superconducting/superfluid state, further decoupling the
components. Recently, there has been a number or relevant papers
aiming at improving the description and understanding of non trivial
interactions between the various particle species.
\citet*{2011MNRAS.410..805G} derived the MHD equations for superfluid
and type II superconducting neutron stars, by using a Newtonian
variational formalism and clarifying several aspects of these systems.
Their equations were consistent with those determined by
\citet{1991AnPhy.205..110M} and \citet{1991ApJ...380..515M,
  1991ApJ...380..530M}, and they recognized the role of the London
magnetic field in the superconducting equivalent to Amp\'er's
law. This formalism has been later applied by \citet{Graber2015} and
\citet{Elfritz2016} to study the induction equation of superconducting
neutron stars and to estimate the (very long) timescales of the
magnetic field evolution. More recently, \citet{2016PhRvD..93f4033G}
and \citet{2016PhRvD..94h3006G} have derived and further extended the
HVBK formalism originally derived to describe the dynamics of
superfluid helium \citep{HV56, H60,BK61,Kbook} to general
relativity. This formalism allows to study superfluid and type I and
type II superconducting neutron stars with thermal effects. The
correct implementation of the buoyancy effect has also been discussed
in \citet{2017arXiv170106870D}. 

Generally the various formalisms provide equivalent descriptions of
the physical system, and they agree on the most part of the equations,
but there are also discrepancies.  The main controversy, which
appeared recently, concerns the form of the induction equation in
superconducting neutron stars.  The electric field determined by
\citet{2016PhRvD..94h3006G} has an extra contribution of the fluxtube
tension which is absent in the derivation of
\citet{2011MNRAS.410..805G} and \citet{Graber2015}. As further
emphasized in \citet{2017arXiv170106870D}, this leads to a difference
of several orders of magnitude in the estimated evolution timescales.

The paradox is that for type II superconducting stars the equations,
which have been derived by these two groups, are completely equivalent
with the only exception of the induction equation. This issue, as we
will show later, is due to the different assumptions the authors have
made to find the electric field, in particular, neglecting the
inertial terms in the momentum and mass conservation equations to
estimate the long-term timescales.  In this paper we revisit these
assumptions, discuss their limitations, and propose how to improve the
current calculations and to reconcile apparently opposed results.  We
will proceed gradually, from the simplest case to the most complex,
showing that the basic math and physics assumptions are analogous in a
simple case and a complex superconducting liquid.

The paper is organized as follows. In Section \ref{sec:scheme} we
describe the case of normal matter. We start by considering an
electron-proton plasma, excluding neutrons for clarity, and then go on
to discuss the more general case when neutrons are included as
well. The superconducting case is treated in Section
\ref{sec:x}. Finally, Section \ref{sec:con} is dedicated to the
concluding remarks.

\section{Magnetic field evolution in normal matter} \label{sec:scheme}
\subsection{Magnetized two-component plasma} 

We begin by studying the simplest case. Consider the dynamical
equations for a two-fluid system composed of two charged components.
These two components are denoted with the letters p (positively
charged with charge $+e$) and e (negatively charged with charge $-e$).
For simplicity, we also assume that each component individually obeys
a barotropic equation of state. In addition to the number (and mass)
conservation for each fluid,
\begin{equation}
\frac{ \partial n_{\x} }{\partial t } + \mvb{\nabla} \cdot \left(
n_{\x} \mvb{v}_{\x} \right) = 0
\label{eq:mc} \, ,
\end{equation}
the Euler equations for the two species can be written in the
following form,
\begin{align}
\rho_{\p}\left(\frac{\partial \mvb{v}_{\p}}{\partial t}
+ \mvb{v}_{\p} \cdot \mvb{\nabla} \mvb{v}_{\p}\right)
& + n_{\p} \mvb{\nabla} \mu_{\p} + \rho_{\p} \mvb{\nabla} \Phi = \nn \\
& = \mvb{F}_{\rm pe}
+ e n_{\p} \left( \mvb{E} + \frac{\mvb{v}_{\p}}{c} \times \mvb{B} \right) 
\label{eq:va4} \, , \\
\rho_{\e}\left(\frac{\partial \mvb{v}_{\e}}{\partial t}
+ \mvb{v}_{\e} \cdot \mvb{\nabla} \mvb{v}_{\e}\right)
& + n_{\e} \mvb{\nabla} \mu_{\e} + \rho_{\e} \mvb{\nabla} \Phi = \nn \\
& = \mvb{F}_{\rm ep}
- e n_{\e} \left( \mvb{E} + \frac{\mvb{v}_{\e}}{c} \times \mvb{B} \right)
\label{eq:vb4}  \, . 
\end{align}
Here $\rho_\x$, $n_\x$, $\mu_\x$, and $\mvb{v}_\x$ are, respectively,
the mass density, number density, chemical potential and velocity of
the x-fluid ($\rm x=e,p$), $\mvb{E}$ and $\mvb{B}$ are the electric
and magnetic fields measured in the laboratory frame, and $\Phi$ is
the gravitational potential. The quantity $\mvb{F}_{\x\y}$ is a drag
force between the two fluids of the form $\mvb{F}_{\x\y} = -
\frac{m_{\x} n_{\x}}{\tau_{\x\y} } ( \mvb{v}_{\x} - \mvb{v}_{\y}) = -
\mvb{F}_{\y\x}$, with $\tau_{\x\y}$ being relaxation times (note that
$\rm x\neq y $).

Adding equations (\ref{eq:va4}) and (\ref{eq:vb4}) one obtains an
equation similar to the Euler equation for a magnetized
fluid. Assuming local charge neutrality ($n_{\p} = n_{\e} \equiv
n_{\rm c}$) and dropping the inertial terms, this reduces to the
magnetohydrostatic equilibrium equation
\begin{equation}
n_{\rm c} \mvb{\nabla} \left( \mu_{\p} + \mu_{\e} \right)
+ \rho_{\rm c} \mvb{\nabla} \Phi = \mvb{F}_{\rm L}
\label{eq:vab} \, ,  
\end{equation}
where $\rho_{\rm c} = \rho_{\p} + \rho_{\e} = (m_{\p} + m_{\e}) n_{\rm
  c}$ is the total mass density of the charged fluid and the Lorentz
force is given by
\begin{equation}
\mvb{F}_{\rm L} = \frac{\mvb{j}\times\mvb{B}} {c}
\label{eq:FL} \,.
\end{equation}
The current $\mvb{j}$ is related to the relative velocity between the
two charged fluids $\mvb{u} \equiv \mvb{v}_{\p} - \mvb{v}_{\e}$
through $\mvb{j} = e n_{\rm c} \mvb{u}$.  To close the system one ha
to solve the Poisson equation to determine $\Phi$ and can specify the
current in terms of the magnetic field by Amp\`ere's law,
\begin{equation}
\mvb{j} = \frac{c}{4\pi} \mvb{\nabla} \times \mvb{B} 
\label{eq:Amp} \, . 
\end{equation}

We can combine the chemical potentials and
the gravitational terms in equation (\ref{eq:vab}) into a single
gradient, which then implies that the Lorentz force per unit charge,
$\mvb{F}_{\rm L}/n_{\rm c}$ must be the gradient of a scalar
function. This requirement can be expressed as
\begin{align}
\nabla \times \left(\frac{\mvb{F}_{\rm L}}{{n_{\rm c}}}\right) = 0~.
\label{eq:gradient}
\end{align}
This is an important point to bear in mind when we consider the
electric field from the dynamical equations below.

The evolution of the magnetic field is given by the induction
equation,
\begin{equation}
\frac{\partial \mvb{B} }{\partial t} 
= - c \mvb{\nabla} \times \mvb{E} \, .
\label{eq:ind0}
\end{equation}
To study the long-term evolution of the magnetic field it is necessary
to adopt some approximations in order to isolate the effects of the
slow processes from the fast dynamics. A common practice is to neglect
the inertial terms in writing the electric field from the Euler
equations. This means that velocities are assumed to be small and to
vary on much longer timescales than any of the relaxation times. In
that limit, the electric field can be written from either of the two
equations (\ref{eq:va4}) and (\ref{eq:vb4}), as
\begin{align}
\mvb{E}{|_\p} & \simeq - \frac{\mvb{F}_{\rm L}}{e n_{\rm c}}
- \frac{\mvb{F}_{\rm pe}}{e n_{\rm c}} - \frac{\mvb{v}_{\e}}{c} \times \mvb{B}
+ \frac{1}{e} \mvb{\nabla} \mu_{\p} + \frac{m_{\p}}{e} \mvb{\nabla} \Phi
\label{eq:va2} \, , \\
\mvb{E}{|_\e} & \simeq - \frac{\mvb{F}_{\rm pe}}{e n_{\rm c}} -
\frac{\mvb{v}_{\e}}{c} \times \mvb{B} - \frac{1}{e} \mvb{\nabla} \mu_{\e}
- \frac{m_{\e}}{e} \mvb{\nabla} \Phi
\label{eq:vb2} \, , 
\end{align}
where the label of the electric field specifies the Euler equation
from which it has been derived. Here, the first equation is written in
a way showing explicitly that the difference between the two
expressions is the Lorentz force per unit charge and some gradient
terms. Obviously, the two expressions are exactly equivalent if one
uses the equilibrium equation (\ref{eq:vab}). As noted above, however,
this equilibrium equation additionally requires the Lorentz force per
unit charge to be a gradient, imposing a severe restriction on the
form of the magnetic field which is not guaranteed to be automatically
satisfied as the magnetic field evolves. Thus, the two equations are
equivalent for the induction equation only if the magnetic field
further satisfies the restriction in equation (\ref{eq:gradient}).

More generally, when $\mvb{F}_{\rm L}/{n_{\rm c}}$ has a solenoidal
part in addition to the irrotational part, it no longer satisfies the
static equilibrium equation (\ref{eq:vab}), and velocity fields
\emph{must} be present. This point is occasionally overlooked. In this
case, the approximate forms of $\mvb{E}|_\p$ and $ \mvb{E}|_\e$ from
the above equations will no longer lead to the same magnetic field
evolution. In order to see which of the two approximations is better,
we can write the exact vector $\mvb{E}$ in the following two forms
\begin{align}
\mvb{E} & = \mvb{E}|_\p+ \frac{m_{\p}}{e} \left( \frac{\partial \mvb{v}_{\p}}{\partial t}
+ \mvb{v}_{\p} \cdot \mvb{\nabla} \mvb{v}_{\p} \right)
\label{eq:va3} \, , \\
\mvb{E} & =  \mvb{E}|_\e - \frac{m_{\e}}{e}  \left( \frac{\partial \mvb{v}_{\e}}{\partial t}
+ \mvb{v}_{\e} \cdot \mvb{\nabla} \mvb{v}_{\e} \right)
\label{eq:vb3} \, , 
\end{align}
where we have re-introduced the inertial terms. We have two other
constraints: (i) the mass carried by electrons is much smaller than
that of the protons, $m_{\e} \ll m_{\p}$, and (ii) the two fluids have
very similar velocities, $\mvb{v}_{\p}\simeq \mvb{v}_{\e}$, i.e.\ the
macroscopic electric current is very small compared to the mass
current because of the strong electromagnetic coupling between the
opposed charges. If both conditions apply, it is more reasonable to
assume that $\mvb{E}|_\e$ is a better approximation of the electric
field, as the correction due to the inertial terms of the electrons is
smaller.

\subsection{Generalized Ohm's law in neutron star matter} \label{sec:gOhm}

More generally, let's consider the case when neutrons are also present
and interact with protons and electrons. We now have equations of
motion for three constituents, with their respective frictional
couplings
\begin{align}
\rho_{\p} \left( \frac{\partial \mvb{v}_{\p}}{\partial t}
+ \mvb{v}_{\p} \cdot \mvb{\nabla} \mvb{v}_{\p} \right) & + n_{\p} \mvb{\nabla} \mu_{\p}
+ \rho_{\p} \mvb{\nabla} \Phi = \nn \\ 
= & \ e n_{\p} \left( \mvb{E} + \frac{\mvb{v}_{\p}}{c} \times \mvb{B} \right) 
+ \mvb{F}_{\rm pe} + \mvb{F}_{\rm pn} 
\label{eq:va5} \, , \\
\rho_{\e}   \left( \frac{\partial \mvb{v}_{\e}}{\partial t}
+ \mvb{v}_{\e} \cdot \mvb{\nabla} \mvb{v}_{\e} \right) & + n_{\e} \mvb{\nabla} \mu_{\e}
+ \rho_{\e} \mvb{\nabla} \Phi =  \nn \\
= & - e n_{\e} \left( \mvb{E} + \frac{\mvb{v}_{\e}}{c} \times \mvb{B} \right)
+ \mvb{F}_{\rm ep} + \mvb{F}_{\rm en} 
\label{eq:vb5}  \, , \\
\rho_{\n} \left( \frac{\partial \mvb{v}_{\n}}{\partial t}
+ \mvb{v}_{\n} \cdot \mvb{\nabla} \mvb{v}_{\n} \right) & + n_{\n} \mvb{\nabla} \mu_{\n}
+ \rho_{\n} \mvb{\nabla} \Phi = \mvb{F}_{\rm np}  + \mvb{F}_{\rm ne}  
\label{eq:vc5} \, .
\end{align}

Adding the three equations, assuming local charge neutrality, and
dropping the inertial terms, one obtains the global static equilibrium
equation
\begin{equation}
n_{\rm c} \mvb{\nabla} \left( \mu_{\p} + \mu_{\e} \right) + n_{\n} \mvb{\nabla}  \mu_{\n}
+ \rho \mvb{\nabla} \Phi = \mvb{F}_{\rm L}
\label{eq:eqMtot} \, ,  
\end{equation}
where $\rho = \rho_{\p} + \rho_{\e} + \rho_{\n}$ is the total mass
density.  This equation can also be written in terms of the pressure
gradient, since $\mvb{\nabla} P = n_{\p}\mvb{\nabla}\mu_{\p} +
n_{\e}\mvb{\nabla}\mu_{\e} + n_{\n}\mvb{\nabla}\mu_{\n}$.  In this
case, for a barotropic equation of state, magnetostatic equilibrium
now requires that the Lorentz force per unit mass $\mvb{F}_{\rm
  L}/\rho$ be a gradient. This leads to the well-known Grad--Shafranov
equation, which determines the structure of the magnetic field.

A standard way to obtain the electric field to study the dynamics of a
multi-constituent plasma is to consider the appropriate linear
combinations of equations (\ref{eq:va5}) and (\ref{eq:vb5}) to derive
a generalized Ohm's law describing the time variation of the electric
current \citep[see e.g.][]{2004prma.book.....G}.  The same procedure
has been applied to neutron stars in different works
\citep{1992ApJ...395..250G, 1995MNRAS.273..643S, 2017MNRAS.465.3416P}
to obtain a general expression for the electric field. Assuming
$m_{\p} \gg m_{\e}$, neglecting inertial terms, and dropping gradients
which are inconsequential for the induction equation (\ref{eq:ind0}),
we get
\begin{align}
\mvb{E} \simeq \frac{\mvb{j}}{\sigma_0} - \frac{\mvb{v}_{\rm c}}{c} \times \mvb{B} 
+ \frac{\mvb{j}}{e n_{\rm c} c } \times \mvb{B} 
\,.
\label{eq:gO2}
\end{align}
Here $\sigma_0$ denotes the electrical conductivity in the absence of
a magnetic field, and the velocity of the charged fluid is given by
\begin{align}
\mvb{v}_{\rm c} = \frac{\rho_{\p}\mvb{v}_{\p} + \rho_{\e}\mvb{v}_{\e}
} {\rho_{\rm c}} \, .
\end{align}

We note that the $\mvb{v}_{\rm c} \times \mvb{B}$ and $\mvb{j} \times
\mvb{B}$ terms in the previous equation can be written in a number of
different, but equivalent, ways. The three basic unknowns are the
three velocities of the constituents, and in principle, any linear
combination of them can be used. A natural choice for one of the
velocities is the current (or equivalently $\mvb{u}$, the relative
velocity between protons and electrons), which gives the last (Hall)
term. We have decided to use the charged plasma velocity $\mvb{v}_{\rm
  c}$ to describe the second term, usually identified with ambipolar
diffusion since it physically represents the advection of magnetic
field lines with the charges, which modifies the magnetic field even
in the limit when electron and proton velocities are the same (and no
currents are present). The third velocity could be, for example, the
velocity of neutrons $\mvb{v}_{\n}$, or the hydrodynamical velocity of
the fluid
\begin{align}
\mvb{v} = \frac{\rho_{\p}\mvb{v}_{\p} + \rho_{\e}\mvb{v}_{\e} +
  \rho_{\n}\mvb{v}_{\n}} {\rho} \, ,
\end{align}
which are usually neglected when the background is assumed to be in
equilibrium.  We also note that Equation (\ref{eq:gO2}) is consistent
with the electric field obtained from the electron Euler equation
(equation \ref{eq:vb2}), in the limit $m_{\e} \ll m_{\p}$.

\subsection{Long-term evolution of the magnetic field} \label{sec:amf}

A common approach to study the effects of the slow processes is to
consider a system in steady state and filter out the short timescale
dynamics (e.g. sound and Alfv\'{e}n waves). This is done by neglecting
the inertial terms in equations
(\ref{eq:va5})-(\ref{eq:vc5}). However, as noted before, for a
barotropic fluid, equation (\ref{eq:eqMtot}) then requires that the
Lorentz force per unit mass be a gradient. Even if the fluid is not
barotropic, in axisymmetry, the hydrostatic terms cannot balance
arbitrary magnetic fields, particularly the $\phi$ component. This
would inevitably result in the appearance of accelerations and the
activation of fast dynamical flows.  Therefore, the configuration of
the magnetic field would rapidly change before the Hall and Ohmic
processes had time to further influence the evolution.

This problem does not arise in the neutron star crust. Since it is
solid, a matter flow cannot be established and any additional force
can be balanced by the elastic response of the crust (up to the
breaking point). However, in the fluid core, the quasi-static
evolution shows serious limitations. Assuming a slow evolution is
still a useful compromise to investigate the effects of the Hall drift
and Ohmic dissipation, but it remains unclear what the magnetic field
geometry will be after the fluid core is relaxed by fast dynamical
processes.

In general, allowing for fluid motion and acceleration, and using the
equations of continuity (equation \ref{eq:mc}), the total momentum
equation can be written as
\begin{align}
\frac{\partial (\rho \mvb{v})}{\partial t} + \mvb{\nabla} \cdot \left( \rho\mvb{v}\mvb{v} 
+ \frac{\rho_{\rm c}\rho_{\n}}{\rho}\mvb{w}\mvb{w} + \frac{\rho_{\e}\rho_{\p}}{\rho_{\rm c}}\mvb{u}\mvb{u}\right) & \nn \\
+ \mvb{\nabla} P + \rho\mvb{\nabla}\Phi
& = \mvb{F}_{\rm L}
\label{eq:eqM2} \, . 
\end{align}
where we have defined $\mvb{w} = \mvb{v}_{\rm c} - \mvb{v}_{\n}$.  For
stationary flows the partial derivatives $\partial/\partial t$ are
neglected, and if one can further argue that the $\mvb{w}$ and
$\mvb{u}$ terms are small compared to the $\mvb{v}$ term, then this
equation reduces to the one fluid case.

Although an evident improvement over the static assumption, the
inclusion of velocities in the stationary regime must be accompanied by a new
advective term in the induction equation, which may become dominant.
Equation (\ref{eq:eqM2}) is not easy to solve in the general case. As
far as we know, in the literature there are not yet numerical
solutions which describe the internal magnetic field of neutron stars
with flow motion. Analytical solutions have been presented only for
simplified cases
\citep{1956ApJ...124..232C,1981ApJ...245..764T,1982ApJ...252..775T}.

\section{Magnetic field evolution in a superconducting neutron star core} \label{sec:x}

Protons in the core of a neutron star undergo a phase transition to a
type-II superconducting state when the temperature drops below
$\lesssim 10^{10}$ K. The magnetic field then penetrates the core in
the form of a dense array of thin fluxtubes.  In this section, we
consider the dynamical equations by using the Newtonian HVBK
formalism\footnote{We use Newtonian theory in order to have an easier
  comparison with the formalism of \citet{2011MNRAS.410..805G}, but
  our discussion is equally valid in the relativistic limit.} for a
plasma formed by normal electrons, superconducting protons and
superfluid neutrons.  These particles are respectively indicated with
the letters e, p and n.  For simplicity, we neglect the entrainment
between nucleons and we omit thermal fluctuations in the
superconducting protons and superfluid neutrons, i.e. we are well
below the critical temperatures.  All protons are therefore in the
superconducting state and quasi-particle excitations are absent.  We
consider a locally neutral system, $n_{\p} = n_{\e} \equiv n_{\rm c}$.

As in the normal case, the dynamics of this plasma at zero
temperature can be described by a system of three Euler and three mass
conservation equations. It is possible to choose the Euler and mass
conservation equations for each fluid species, or otherwise use a
combination of these three equations to describe the motion of the total
fluid. The HVBK formalism provides the total conservation equation and
describes separately the Euler equations for the superconducting and superfluid 
particles \citep{1991AnPhy.205..110M, 2016PhRvD..94h3006G}.

The hydrodynamical system of equations formed by a ``normal'' and a
``superconducting'' component are very similar to the normal fluid
case detailed in the previous section, with a few remarkable
differences. In particular, we note that:
\begin{enumerate}

\item The true Lorentz force $\mvb{F}_{\rm L}$ is negligible, because,
  in type II superconductors, Amp\`ere's law connects the London field
  $\mvb{b}_{\rm L} $ to the macroscopic average currents
  \citep{2011MNRAS.410..805G},
\begin{equation}
\mvb{\nabla} \times \mvb{b}_{\rm L} = \frac{4\pi}{c} \mvb{j} \, .  \label{eq:Amp}
\end{equation}
and the London field is very weak $\mvb{b}_{\rm L} \ll \mvb{B}$.  For
the same reason, at the hydrodynamical averaged scale it is safe to
assume that $u \ll v_{\p} \approx v_{\e}$ (i.e.\ no macroscopic
currents).

\item
However, there is now a new superconducting force, the
tension/buoyancy of the fluxoids $\mvb{T}$, which can be written in
the following compact form,
\begin{equation}
\mvb{T} = \frac{\mvb{\mathcal{C}} \times \mvb{B} }{c}\, ,
\end{equation}
where
\begin{equation}
\mvb{\mathcal{C}} \equiv \frac{c}{4\pi} \mvb{\nabla} \times \left( H_{c1} \mvb{\hat{b}} \right)
\, .  \label{eq:Cdef}
\end{equation}
As shown by \citet{2017arXiv170106870D}, the vector field $\mvb{T}$
also contains the contribution of the buoyancy force. Equation
(\ref{eq:Cdef}) has the same mathematical form as Amp\`ere's law in
the normal case, with the difference that the magnetic induction
$\mvb{B}$ is replaced by the vector $H_{c1} \mvb{\hat b}$, where
$H_{c1}$ is the lower critical magnetic field \citep{Tinkham2004}.  In
type II superconducting neutron stars,
\begin{equation}
H_{c1} \approx 10^{15} \left(\frac{n_\p}{0.01 \,  \textrm{fm}^{-3}}\right)~{\mathrm G}.
\end{equation}
Note that this is equivalent to replacing the macroscopic current
$\mvb{j}$ in the Lorentz force by an effective current
$\mvb{\mathcal{C}}$.

\item
The mutual friction terms look different and they have a
  different physical origin.  For example, $\mvb{F}_{\p\e}$ describing
  the mutual friction between the protons and electrons is 
\citep{BK61,1991ApJ...380..530M}
\begin{equation}
\mvb{F}_{\p\e} = - \left( 1 + \alpha \rho_{\p}\right) \, \mvb{T} -
\beta \rho_{\p} \, \mvb{\hat{b}} \times \mvb{T}
- \gamma \rho_{\p} \mvb{\hat{b}} \left( \mvb{B} \cdot \frac{\mvb{\mathcal{C}}}{c} \right)
\, . \label{eq:Fmf}
\end{equation}
The third term is normally neglected because its strength is small
compared to the other two \citep{BK61}. The coefficients $\alpha$ and
$\beta$ are related to the dimensionless drag coefficient $R$
(typically $R\sim 10^{-4}$) by \citep{2017arXiv170106870D}
\begin{equation}
 \alpha \rho_{\p} = - \frac{1}{1+R^2} \, \qquad \mbox{and} \qquad 
\beta \rho_{\p} = \frac{R}{1+R^2} \, .
\end{equation}
For brevity, we do not discuss the particular forms of the vector
fields $\mvb{F}_{\e\n}$ and $\mvb{F}_{\p\n}$, describing mutual
friction forces between neutrons and the charged particles, as these
are not important for this work.  For example, they can describe the
electron scattering off the magnetized core of a neutron vortex
\citep{ALS84} or the interaction between proton and neutron vortices
\citep[see e.g.][]{1989ASIC..262..457S,1998ApJ...492..267R}.

\end{enumerate}

We now consider the Euler equations.  After neglecting inertial terms,
and other terms of the order of the weak London field, the equations 
for the superconducting protons, normal electrons and superfluid
neutrons in the laboratory reference frame read 
\begin{align}
n_{\p} \mvb{\nabla} \hat{\mu}_{\p}  + \rho_{\p} \mvb{\nabla} \Phi = & 
\mvb{T} + \mvb{F}_{\p \e}  +  \mvb{F}_{\p \n} + e n_{\p} \left( \mvb{E} + \frac{\mvb{v}_{\p}}{c} \times \mvb{B}\right)
\label{eq:vp} \, , \\
n_{\e} \mvb{\nabla} \mu_{\e} + \rho_{\e} \mvb{\nabla} \Phi = & 
- \mvb{F}_{\p\e} + \mvb{F}_{\e\n} - e n_{\e} \left( \mvb{E} + \frac{\mvb{v}_{\e}}{c} \times \mvb{B} \right) \, , \label{eq:ve} \\
 n_{\n} \mvb{\nabla} \hat \mu_{\n} + \rho_{\n} \mvb{\nabla} \Phi = &  - \mvb{F}_{\e\n} -  \mvb{F}_{\p\n}    \, .  \label{eq:vn} 
\end{align}
Here, the chemical potentials $\hat{\mu}_{\p}$ and $\hat{\mu}_{\n}$
may also contain contributions due to the magnetic field energy
\citep{2011MNRAS.410..805G} and kinetic terms.  Equation (\ref{eq:ve})
is equivalent to the electron Euler equation provided in
\citet{2011MNRAS.410..805G}, since the mutual friction force density
(\ref{eq:Fmf}) is exactly the force used in their work and in
\citet{Graber2015}.  In equation (\ref{eq:vp}) we have replaced the
force $f_{\p}$ which was used in \citet{2016PhRvD..94h3006G}, with the
relation $\rho_{\p} n_{\p} f_{\p} = - \mvb{T} - \mvb{F}_{\p\e}$.

The sum of equations (\ref{eq:vp})-(\ref{eq:vn}) gives
\begin{equation}
n_{c} \mvb{\nabla} \left( \hat{\mu}_{\p}  + \mu_{\e} \right) + n_{n} \mvb{\nabla} \hat{\mu}_{\n} + \rho
\mvb{\nabla} \Phi = \mvb{T}  \, ,
\label{eq:vtot2}
\end{equation}
which implies that, in hydrostatic equilibrium, and for a barotropic
equation of state
\begin{equation}
\mvb{\nabla} \times \left( \frac{\mvb T}{\rho} \right) = 0 \, . 
\label{eq:vpe}
\end{equation}
This is the condition for magnetostatic equilibrium in superconducting
neutron stars, and has been studied in a number of works
\citep{Roberts81, 2008MNRAS.383.1551A, 2013PhRvL.110g1101L,
  2013MNRAS.431.2986H, 2014MNRAS.437..424L, 2015MNRAS.452.3246P}.

\subsection{Evolution timescales} \label{sec:gOhmsuper}

Surprisingly, although the momentum equations given by
\cite{2011MNRAS.410..805G} and \cite{2016PhRvD..94h3006G} are
equivalent, \citet{2017arXiv170106870D} find that their evolution
timescales differ by several orders of magnitude from the results of
\cite{2011MNRAS.410..805G}.  We now discuss the cause of this
discrepancy.  In \citet{2016PhRvD..94h3006G} and
\cite{2017arXiv170106870D}, their approximation for the electric field
is derived from the Euler equation of the superconducting protons
(equation \ref{eq:vp}),
\begin{equation}
e \mvb{E} |_{\p} \equiv - \frac{\mvb{T} + \mvb{F}_{\p\n}}{n_c} -
\left( \frac{e}{c} \mvb{v}_{\p} \times \mvb{B} + \frac{\mvb{F}_{\rm
    pe}}{n_{\rm c}} \right) + \mvb{\nabla} \hat{\mu}_{\p} + m_{\p} \mvb{\nabla} \Phi \, ,  
\label{eq:Eps}
\end{equation}
which, using equation (\ref{eq:vn}), can be written as
\begin{align}
e \mvb{E} |_{\p} \equiv - \frac{\mvb{T}}{n_c} + \frac{\mvb{F}_{\e\n}}{n_c} -
\left( \frac{e}{c} \mvb{v}_{\p} \times \mvb{B} + \frac{\mvb{F}_{\rm
    pe}}{n_{\rm c}} \right)  \nn \\ 
    + \frac{ n_c \mvb{\nabla} \hat{\mu}_{\p} +  n_n \mvb{\nabla} \hat{\mu}_{\n} + (\rho_{\p}+\rho_{\n}) \mvb{\nabla} \Phi }{n_c} \,
, \label{eq:Et}
\end{align}
In \citet{2011MNRAS.410..805G}, it is determined from the
electron equation (\ref{eq:ve}),
\begin{equation}
e \mvb{E} |_{\e} \equiv \frac{\mvb{F}_{\e\n}}{n_c} - \left(
\frac{e}{c} \mvb{v}_{\e} \times \mvb{B} + \frac{\mvb{F}_{\rm pe}
}{n_{\rm c}} \right) - \mvb{\nabla} \mu_{\e} - m_{\e} \mvb{\nabla} \Phi\,
. \label{eq:EGAS}
\end{equation}
In both cases the inertial terms are neglected. 

It is evident that both approximations to the electric field are
equivalent, because one can use equation (\ref{eq:vtot2}) to change
from one expression to the other.  The problem here is that the
apparently large contribution of the tension term $\mvb{T}/n_{\rm c}$,
estimated from equation (\ref{eq:Eps}), in reality must be considered
in combination with the total momentum equation. In fact, from
equation (\ref{eq:Et}), we can see that the various gradient terms
balance the contribution of the $\mvb{T}/n_{\rm c}$.

To highlight the issue, let us momentarily proceed by ignoring the
gradient terms in equations (\ref{eq:Eps}) and (\ref{eq:EGAS}) as well
as ${\mvb{F}_{\e\n}}$ and ${\mvb{F}_{\p\n}}$ in comparison with
${\mvb{F}_{\p\e}}$ \citep[see][]{2011MNRAS.410..805G}. As noted above,
this procedure may be misleading when the total momentum equation is
not considered.  After using the explicit form of the mutual friction
force (\ref{eq:Fmf}), the electric field approximated by equations
(\ref{eq:Eps}) and (\ref{eq:EGAS}) can be written in the following
compact form
\begin{equation}
\mvb{E} = - \frac{\mvb{v}_{\p}}{c} \times \mvb{B} + \frac{1}{e n_{\rm
    c}} \left( \frac{R^2}{1+R^2} - \chi \right) \mvb{T} + \frac{1}{e
  n_{\rm c}} \frac{R}{1+R^2} \, \mvb{\hat b} \times \mvb{T} + \dots \,
, \label{eq:gSO3}
\end{equation}
where the parameter $\chi$ provides the link between the two cases:
$\chi=1$ for equation (\ref{eq:Eps}), $\chi=0$ for equation
(\ref{eq:EGAS}).  It is important to remark that this parameter
appears only in the second term of equation (\ref{eq:gSO3}), which is
proportional to the fluxtube tension. Comparing with the normal matter
case where (equation \ref{eq:gO2})
\begin{align}
\mvb{E} \simeq \frac{\mvb{j}}{\sigma_0} - \frac{\mvb{v}_{\rm c}}{c}
\times \mvb{B} + \frac{\mvb{F}_{\rm L}}{e n_{\rm c} } \,,
\end{align}
we can note three differences: 1) there is no Ohmic dissipation in the
superconductor; 2) the non-dissipative Hall term ($\propto
\mvb{F}_{\rm L}$) is replaced by a similar term ($\propto \mvb{T}$);
3) there is a new term, proportional to $\mvb{\hat b} \times \mvb{T}$,
which has the same mathematical structure as ambipolar diffusion as
defined in \cite{1992ApJ...395..250G}.

From equation (\ref{eq:gSO3}) and the induction equation, we can
extract two timescales describing the long-term evolution of the
magnetic field. As shown by \citet{Graber2015} and
\citet{2017arXiv170106870D}, one can determine a conservative
(mathematically, a Hall-like term) and a dissipative timescale. The
controversy is about the estimate of the conservative timescale
\begin{equation}
\tau_{\rm con} = \frac{e \rho_p }{m_{\p} c} \frac{4\pi L^2}{H_{\rm c1}} \left(
\frac{R^2}{1+R^2} - \chi \right)^{-1} \,  ,
\end{equation}
with the parameter $L$ denoting a typical length-scale of the system.
To recover the result of \citet{Graber2015}, we take $\chi = 0$, so
that $\tau_{\rm con}= {\cal O}{(R^2)}$, and
\begin{equation}
\tau_{\rm con} \approx 1.3 \times 10^{15} L_6 ^2 ~~\textrm{yr} \,
, \label{eq:tauGr}
\end{equation}
while for \citet{2017arXiv170106870D}, $\chi = 1$, $\tau_{\rm
  con}= {\cal O}{(1)}$ and
\begin{equation}
\tau_{\rm con} \approx 2 \times 10^{8} L_6 ^2 \, ~\textrm{yr} .
\end{equation}

The large difference between the two estimates is apparently caused by
the presence of a much larger term $\propto \mvb{T}/n_{\rm c}$. As
pointed out before, from the total momentum equation (\ref{eq:vtot2}),
we can see that the combination of $ \mvb{T}/n_{\rm c}$ with the
various gradients present in (\ref{eq:Et}) is an irrotational field.
Hence, this term does not lead to any change in the magnetic field
after inserting the electric field in the induction equation.

The timescale (\ref{eq:tauGr}) should be considered as a lower limit,
because if the leading term in the electric field is nearly
irrotational $\tau_{\rm con}$ is even longer.  To correctly estimate
the evolution timescale of the magnetic field, we should first
calculate the solenoidal part of the electric field, after combining
all terms, but this requires a detailed prescription of the magnetic
field geometry and to know the velocity field.  For general magnetic
fields, if the system is not required to be strictly in magnetostatic
equilibrium, the motion of the fluid and inertial terms may not be
negligible, and the approximations usually made are questionable. In
principle, there can be dynamical readjustments that modify the
magnetic field geometry much faster than the slow, secular processes.

\section{Conclusions} \label{sec:con}

We have reconsidered the problem of the magnetic field evolution in a
multicomponent plasma, and discussed the approximations normally used
in the context of long-term evolution of magnetic fields in neutron
stars. The standard approach assumes that hydro-magnetic equilibrium
is reached in the fluid core immediately after formation (within tens
of rotation/Alfv\'en timescales). After this first stage, slow
processes, such as Ohmic dissipation, Hall drift and ambipolar
diffusion, modify the magnetic field on much longer timescales. This
approximation is motivated by the multi-scale nature of the problem,
which makes it extremely difficult to carry out a numerical study of
the complete system of MHD equations. However, the assumption of
hydro-magnetic equilibrium also places a strong constraint on the form
of the magnetic force, which must be mostly irrotational (in the
absence of inertial terms). Therefore, terms in the electric field
proportional to the magnetic force (either the Lorentz force for
normal matter, or the fluxoid tension for superconducting matter) must
be treated with extreme caution when estimating evolution timescales,
as only the solenoidal parts contribute to the induction equation, and
these can be many orders of magnitude smaller.

The preceding remark clarifies a recent controversy that appeared in
the literature, about the ``correct'' expression for the induction
equation in superconducting neutron stars. This is an important issue,
as a disagreement of seven orders of magnitude between estimates of
timescales (see \citealt{Graber2015} and
\citealt{2017arXiv170106870D}) would lead to very different
conclusions about the evolutionary scenario. Our analysis shows that
the evolution timescale of \citet{2017arXiv170106870D} is likely an
overestimation, and that the one given in \citet{Graber2015} and
\citet{Elfritz2016} should be closer to describe the slow evolution of
the magnetic field in a neutron star core.

More generally, as the magnetic field evolves, the magnetic force per
unit mass will very likely quickly acquire a solenoidal component. In
the fluid core, this will inevitably activate fluid motions which,
acting on dynamical timescales, would modify the magnetic field faster
than the Ohmic and Hall processes (or their equivalent dissipative and
non-dissipative processes in superconducting cores).  This problem
does not arise for the crust of a neutron star, which is instead
modeled as a solid ion lattice plus a single electron fluid. The
elasticity of the crust can sustain small deviations from equilibrium
for any magnetic field of reasonable strength.

A significant improvement in our understanding of magnetic field
evolution in neutron star cores would be to construct stationary (but
not static) solutions of the momentum and mass conservation equations,
which would allow us to obtain the velocity field explicitly, and to
incorporate the corresponding advective terms in the electric field
and the induction equation. This may significantly alter the secular
evolution timescales in neutron star cores.


\section*{Acknowledgements}
A.P. acknowledges support from the European Union under the Marie
Sklodowska Curie Actions Individual Fellowship, grant agreement
n$^{\rm o}$ 656370.  This work is supported in part by the Spanish
MINECO grant AYA2015-66899-C2-2-P, and by the New
Compstar COST action MP1304.


\nocite*
\bibliographystyle{mn2e}

\label{lastpage}

\end{document}